\documentclass[hidelinks,12pt]{article}
\usepackage{amsmath}
\usepackage{graphicx,psfrag,epsf}
\usepackage{enumerate}
\usepackage[square,numbers]{natbib}
\usepackage{url} 
\usepackage[document]{ragged2e}
\usepackage[backref]{hyperref}
\usepackage{multirow}
\usepackage[table,xcdraw]{xcolor}
\newcommand{\blind}{0}

\usepackage{amssymb}
\usepackage{bm}
\usepackage{amsthm}
\usepackage[titletoc,title]{appendix}
\usepackage{rotating}
\usepackage[skip=0.5\baselineskip]{caption}
\usepackage{tabularx}
\usepackage{longtable}
\newcolumntype{b}{X}
\newcolumntype{s}{>{\hsize=.5\hsize}X}

\newcommand{\mbb}[1]{\mathbb{#1}}
\newcommand{\mbf}[1]{\boldsymbol{#1}}
\newcommand{\mcal}[1]{\mathcal{#1}}
\newcommand{\mrm}[1]{\textrm{#1}}

\theoremstyle{definition}

\makeatletter
\newcommand\incircbin
{%
  \mathpalette\@incircbin
}
\newcommand\@incircbin[2]
{%
  \mathbin%
  {%
    \ooalign{\hidewidth$#1#2$\hidewidth\crcr$#1\bigcirc$}%
  }%
}
\newcommand{\owedge}{\incircbin{\wedge}}
\makeatother

\begin{document}

\def\spacingset#1{\renewcommand{\baselinestretch}%
{#1}\small\normalsize} \spacingset{1}


\if1\blind
{
  \title{\bf Title}
  \author{Author 1\thanks{
    The authors gratefully acknowledge \textit{please remember to list all relevant funding sources in the unblinded version}}\hspace{.2cm}\\
    Department of YYY, University of XXX\\
    and \\
    Author 2 \\
    Department of ZZZ, University of WWW}
  \maketitle
} \fi

\if0\blind
{
  \title{\bf An Association Test Based on Kernel-Based Neural Networks for Complex Genetic Association Analysis}
  \author{Tingting Hou, Chang Jiang and Qing Lu}
  \maketitle
} \fi

\bigskip
\begin{abstract}
The advent of artificial intelligence, especially the progress of deep neural networks, is expected to revolutionize genetic research and offer unprecedented potential to decode the complex relationships between genetic variants and disease phenotypes, which could mark a significant step toward improving our understanding of the disease etiology. While deep neural networks hold great promise for genetic association analysis, limited research has been focused on developing neural-network-based tests to dissect complex genotype-phenotype associations. This complexity arises from the opaque nature of neural networks and the absence of defined limiting distributions. We have previously developed a kernel-based neural network model (KNN) that synergizes the strengths of linear mixed models (LMM) with conventional neural networks. KNN adopts a computationally efficient minimum norm quadratic unbiased estimator (MINQUE) algorithm and uses kernel-based neural network structure to capture the complex relationship between large-scale sequencing data and a disease phenotype of interest. In the KNN framework, we introduce a MINQUE-based test to assess the joint association of genetic variants with the phenotype, which considers non-linear and non-additive effects and follows a mixture of chi-square distributions. We also construct two additional tests to evaluate and interpret linear and non-linear/non-additive genetic effects, including interaction effects. Our simulations show that our method consistently controls the type I error rate under various conditions and achieves greater power than a commonly used sequence kernel association test (SKAT), especially when involving non-linear and interaction effects. When applied to real genetic data from the UK Biobank, our approach identified genes associated with hippocampal volume, which can be further replicated and evaluated for their role in the pathogenesis of Alzheimer's disease.

\end{abstract}

\noindent%
{\it Keywords:}  Human genome, Complex relationships, MINQUE, Deep learning, Hypothesis testing
\vfill

\newpage
\spacingset{1.45} 
\section{Introduction}
\label{sec:intro}
In light of the rapid advancements in high-throughput sequencing technologies, we now possess the capability to harness vast genetic data, marking a stride towards precision medicine informed by human genome findings and other pivotal risk predictors \cite{manolio2009finding}. While Genome-wide association studies (GWAS) have made strides in identifying numerous genetic loci related to complex diseases such as cancer, cardiovascular diseases, and neurodegenerative conditions, these identified variants only account for a small fraction of the heritability \cite{manolio2009finding,Maher2008}. This gap, often referred to as "missing heritability," stems from a range of factors. These include the inherent complexity of the genetic architecture, non-linear genotype-phenotype relationships, potential gene interactions, and the limitations of current genotyping arrays, which might not effectively capture rarer or structural variants \cite{eichler2010missing,phillips2008epistasis}. Additionally, these arrays may not adequately consider shared environmental factors among family members \cite{manolio2009finding}. As we aim to delve deeper into the genetic foundations of various diseases and introduce innovative diagnostic and therapeutic approaches, there's a pressing need for advanced analytical methods. These methodologies, designed to adeptly navigate the intricate genotype-phenotype connections, promise to enhance our understanding of how genomic structures influence diseases, ushering in a new era of personalized medicine \cite{hindorff2009potential}.

Over recent years, the integration of machine learning, particularly deep neural networks, into genomics has been noteworthy \cite{angermueller2016deep}. These networks have revolutionized the field by eliminating the manual extraction of features and skillfully identifying complex patterns in genomic data. Deep learning models are designed to offer multi-layered abstraction, enabling in-depth analysis of genomic sequences and their interactions. While their use is increasing in computational biology areas like functional genomics, they are less mature in domains such as phylogenetics \cite{sapoval2022current}. A significant challenge with neural networks is their "black box" nature, which complicates understanding the biological significance of their outputs \cite{chakraborty2017interpretability}. To tackle this, we introduced a kernel-based neural network model for high-dimensional risk prediction, merging the strengths of linear mixed models (LMM) and traditional neural networks. This model provides a unique framework for deciphering complex genotype-phenotype relationships in high-dimensional data \cite{shen2021kernelbased}, showing superior prediction accuracy with nonlinear activation functions. Expanding on this, we developed a Wald-type test for evaluating the collective impact of multiple genetic variants on a disease phenotype, taking into account non-linear and non-additive effects\cite{hou2023kernel}. Our approach includes two tests: one focusing on linear genetic effects and another exploring complex non-linear and non-additive effects, such as interaction effects. Our method has consistently outperformed the sequence kernel association test (SKAT), especially in scenarios involving non-linear and non-additive effects.

However, this association test faces limitations: 1) It has a substantial computational load. The effectiveness of the estimator's limiting distribution, which is crucial for the test's accuracy, depends on a specific iterative process known as I-MINQUE (Iterative Minimum Norm Quadratic Unbiased Estimation). This process is integral to ensuring that the distribution is functionally independent of the norm choice. However, the iterative nature of I-MINQUE makes the computational process more resource-intensive, potentially limiting the test's applicability in scenarios where computational resources are constrained or when quick analysis is required. 2) conservative type I error under low-dimensional cases.  In cases where the number of genetic variants is substantially smaller than the sample size, the Minque-Type estimators, which are foundational to our Wald-type testing, do not perform as expected. This discrepancy is particularly evident in the type I error rate, where the testing tends to be more conservative than intended. This conservative bias could lead to an underestimation of significant genetic effects, thus impacting the test's overall utility in smaller datasets. It's crucial to acknowledge this limitation, as it suggests that our method, while effective in high-dimensional contexts, might require adjustments or alternative approaches when applied to datasets with fewer genetic variants.

In this paper, we proposed two novel statistics to test the overall genetic effect and linear/nonlinear, or non-additive, genetic effect. These new novel statistics were derived from the MINQUE estimator, which follows the mixture chi-square distribution, ensuring consistent control of the type I error rate across all conditions. Comprehensive simulations were performed to evaluate the improvement of the newly proposed method in accuracy and power over the sequence kernel association test(SKAT) \cite{wu2011rare}. The new testing method was then used to identify the genes associated with hippocampus volume using data from the UK Biobank \cite{sudlow2015uk}.

\section{Methodologies}
\label{sec:meth}
\subsection{Kernel-Based Neural Network}
\begin{figure}[htbp]
\centering
\includegraphics[width=\linewidth]{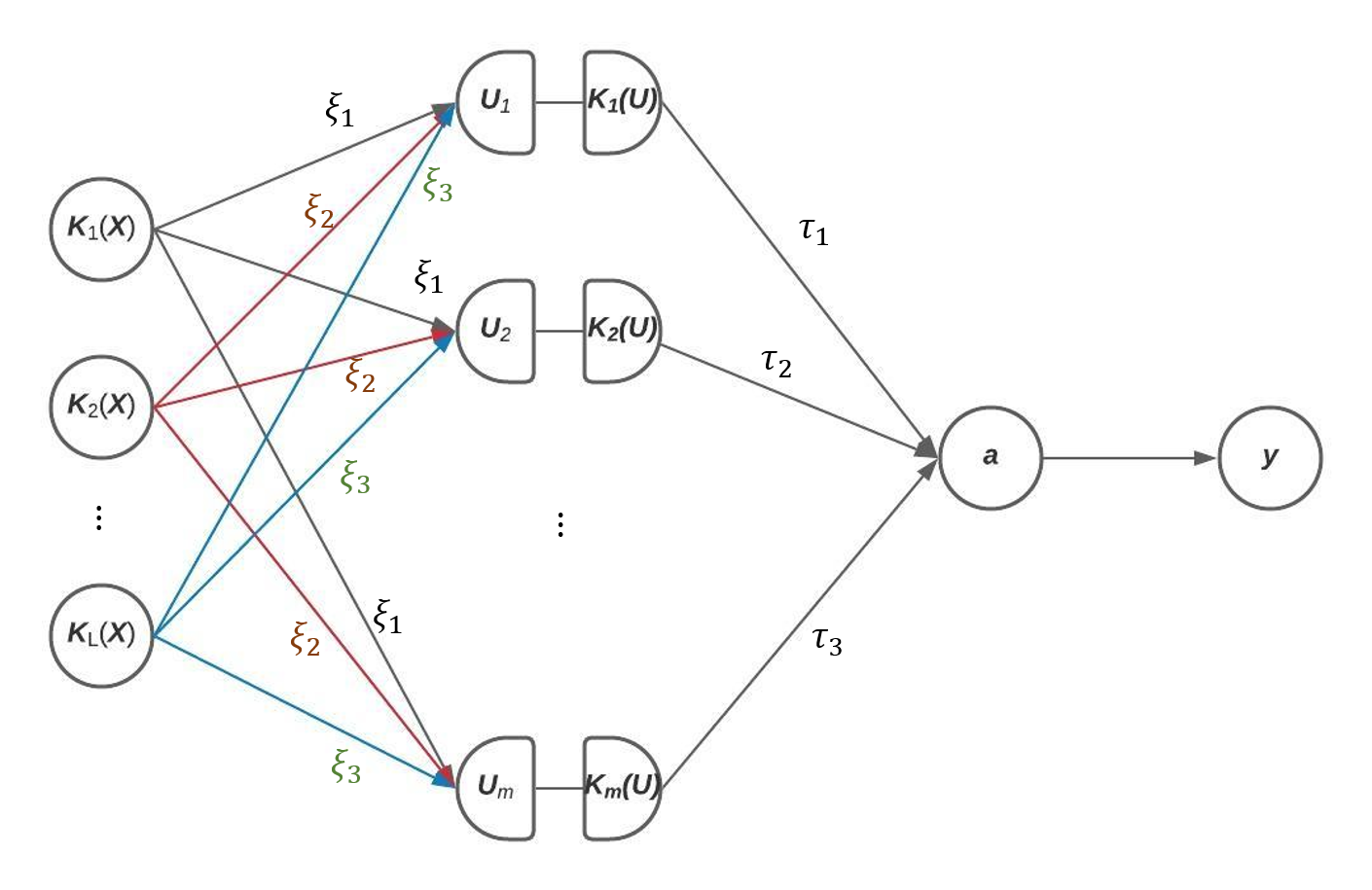}
\caption{An illustration of the hierarchical structure of the kernel-based neural network model.}\label{fig:KNNfig}
\end{figure}

The kernel-based neural network is a promising method that combines the advantages of kernel methods and a mixed linear model, trying to capture the underlying information, and improving the accuracy of genetic risk prediction. Figure \ref{fig:KNNfig} shows a basic hierarchical structure of the kernel-based neural network model. The phenotype  $\bm{y}$ is modeled as random effect model given some hidden variables $\bm{u_1}, \dots, \bm{u_m}$,
\begin{align}
        \bm{y|Z,a} &\sim  \mathcal{N}_n \left(\bm{Z}\bm{\beta}+\bm{a},\phi \bm{I}_n \right)   \\
        \bm{a}|\bm{u_1}, \dots, \bm{u_m} &\sim  \mathcal{N}_n\left(\bm{0},\sum_{j=1}^J \tau_j \bm{K}_j(\bm{U})\right) 
\end{align}
where $n$ is the sample size; $m$ is the number of hidden units in the network; $\bm{Z}$ is design matrix, $\bm{\beta}$ is the vector of fix effects; the covariance matrix of random effect $\bm{a}$ is a positive combination of some latent kernel matrices $\bm{K}_j(\bm{U})= f_j \left[ \frac{1}{m} UU^T\right]$ , where $\bm{U} = [\bm{u_1}, \dots, \bm{u_m}] \in \mathbb{R} ^{n \times m}$. The latent variables $\bm{u_i}$ is constructed from
\begin{align}
    \bm{u_1}, \dots, \bm{u_m} \sim  \mathcal{N}_n\left(\bm{0},\sum_{l=1}^L \xi_l \bm{K}_l(\bm{X})\right) 
\end{align}
where $\bm{K}_l(\bm{X})$, l = 1, . . . , L are kernel matrices constructed based on the genetic variables.  Then the marginal mean and variance of $y$ can be written  as
\begin{align}
\mbb{E}[\mbf{y}] & =\mbb{E}\left(\mbb{E}[\mbf{y}|\mbf{u}_1,\ldots,\mbf{u}_m]\right)=\mbf{0}.\\
\mrm{Var}[\mbf{y}] & =\mbb{E}\left[\mrm{Var}\left(\mbf{y}|\mbf{u}_1,\ldots,\mbf{u}_m\right)\right]+\mrm{Var}\left[\mbb{E}\left(\mbf{y}|\mbf{u}_1,\ldots,\mbf{u}_m\right)\right] \nonumber \\
	& =\sum_{j=1}^J\tau_j\mbb{E}[\mbf{K}_j(\mbf{U})]+\phi\mbf{I}_n\\
	& \simeq\tau f[\mbf{\Sigma}]+\phi\mbf{I}_n,
\end{align}
where $\mbf{\Sigma}=\sum_{l=1}^L\xi_l\mbf{K}_l(\mbf{X})$ (Xiaoxi). If $f(x) = (1+x)^2$, which corresponds to the output polynomial kernel, then $f[\mbf{\Sigma}]=(\mbf{J}+\mbf{\Sigma})^{\owedge2}$, where the symbol $\owedge2$ means the element-wise square. Without loss of generality, in this paper, we consider the case with $L=1$ and $\mbf{K}_1(\mbf{X})=p^{-1}\mbf{XX}^T$. Then, the marginal variance of y can be  written as 
\begin{align}
\mrm{Var}[\mbf{y}] & \simeq\tau f[\mbf{\Sigma}]+\phi\mbf{I}_n \\
	& =\tau\mbf{J}+2\tau\xi_1\frac{1}{p}\mbf{XX}^T+\tau\xi_1^2\frac{1}{p^2}(\mbf{XX}^T)^{\owedge2}+\phi\mbf{I}_n \\
	& =\theta_1 \mbf{J}+\theta_2 \frac{1}{p}\mbf{XX}^T+\theta_3 \frac{1}{p^2}(\mbf{XX}^T)^{\owedge2}+\theta_0 \mbf{I}_n, \label{eq:varKNN}
\end{align}
where the parameters $\theta_0,\theta_1,\theta_2,\theta_3$ can be estimated via the Minimum Norm Quadratic Unbiased Estimator (MINQUE) \cite{rao1970estimation, rao1971estimation,rao1972estimation}.
\clearpage
\subsection{The Asymptotic Distribution of Minque-Type Estimators}

Instead of employing REML \cite{patterson1971recovery} to estimate variance components, we utilize MINQUE \cite{rao1971estimation, rao1970estimation, rao1972estimation}, which has a closed-form solution and can be calculated efficiently. MINQUE uses a quadratic form $\mbf{y}^T\mbf{A}\mbf{y}$ to estimate a linear combination of variance components. The MINQUE matrix $\mbf{A}$ is obtained by minimizing a Euclidean norm of the difference between $\mbf{A}$ and matrix in the quadratic estimator by assuming that we know the random variables in the variance components model. MINQUE provides unbiased estimates for variance components.

Assume $\mbf{y} \sim N(\mbf Z \mbf \beta, \mbf \Sigma)$,
\begin{equation}\label{equ:knncovmatrix}
\mbf{\Sigma}=\sum_{j=0}^J\sigma_j^2 \mbf{K}_j(\mbf{X}).
\end{equation}
The parameters $\mbf{\sigma}=\left(\sigma_0^2,\cdots,\sigma_J^2\right)\in\mbb{R}^J$ can then be estimated via MINQUE. 

Specifically, let $\mbf{V}_j=\mbf{K}_j(\mbf{X}),\ j=0,\cdots,J$,
$\mbf{V}=\sum_{j=0}^J\alpha_j\mbf{K}_j(\mbf{X})$, where  $\mbf \alpha=[\alpha_0^2\dots\alpha_J^2]$ is an initial guess of the true $\mbf \sigma$.
Then we define $$\mbf{P}_v=\mbf{Z}\left(\mbf{Z}^T\mbf{V}^{-1}\mbf{Z}\right)^-\mbf{Z}^T\mbf{V}^{-1}$$ and $$\mbf{Q}_v=\mbf{I}-\mbf{P}_v.$$ The MINQUE estimate of $\mbf \sigma$ is
\begin{equation}\label{equ:mnqvcsestimate}
\mbf{\hat{\sigma}}=\mbf{S}^{-1}\mbf{Q},
\end{equation}
where $\mbf{Q}\in\mbb{R}^J$ is a vector with elements $\mbf{Q}_i=\mbf{y}^T\mbf{Q}_v^T\mbf{V}^{-1}\mbf{V}_i\mbf{V}^{-1}\mbf{Q}_v\mbf{y},\ i=0,\cdots,J$, and $\mbf{S}\in\mbb{R}^{J\times J}$ is a symmetric matrix with elements  $\mbf{S}_{ij}=\mrm{tr}(\mbf{Q}_v^T\mbf{V}^{-1}\mbf{V}_i\mbf{V}^{-1}\mbf{Q}_v\mbf{V}_j), i,j=0,\cdots,J$. In the case of no covariates, $\mbf{P}_v=\mbf{0}$.

We then consider the limiting distribution of an estimator $\hat{\sigma}_i^2 \in \hat{\mbf \sigma}$, which could be represented as
\begin{align}
    \hat{\sigma}_i^2 &= \sum_{j=0}^J c_{ij} \mbf{Q}_i \\
    &=  (\mbf{Q}_v\mbf{y})^T\left(\sum_{j=0}^J  c_{ij} \mbf{V}^{-1}\mbf{V}_i\mbf{V}^{-1}\right) \mbf{Q}_v\mbf{y}\\
    &= \mbf{\epsilon}^T\mbf M_i \mbf{\epsilon} \label{eq:Quadraticforms}
\end{align}
where $c_{ij}$ is the $ij$th element in the matrix $\mbf S^{-1}$, and $\mbf M_i=\sum_{j=0}^J  w_{ij} \mbf{V}^{-1}\mbf{V}_i\mbf{V}^{-1}$, and $\mbf \epsilon \sim N(\mbf 0, \mbf \Sigma)$

Under null hypothesis $\sigma^2_i=0$, we have $\mbf \epsilon \sim N(\mbf 0, \mbf \Sigma_{H_0} )$, where $\mbf \Sigma_{H_0} = \sum_{-i} \sigma^2_j \mbf V_j$, then the statistics $\hat{\sigma}_i^2$ follows a mixture of chi-square distributions\cite{liu2007semiparametric,liu2008estimation}, 
\begin{align}
   \hat{\sigma}_i^2 &=  \left(\mbf \Sigma_{H_0}^{-1/2}  \mbf \epsilon \right)' \mbf \Sigma_{H_0}^{1/2} \mbf M_i \mbf \Sigma_{H_0}^{1/2} \left(\mbf \Sigma_{H_0}^{-1/2} \mbf  \epsilon\right) \\
    &=  \mbf \epsilon^*{}' \Tilde{\mbf M}_i \mbf \epsilon^{*}  \sim \sum_j \lambda_j \chi_1^2
\end{align}
where $\mbf  \epsilon^* = \mbf \Sigma_{H_0}^{-1/2} \mbf  \epsilon, \mbf  \epsilon^* \sim N(\mbf 0, \mbf I)$,  $\Tilde{\mbf M}_i= \mbf \Sigma_{H_0}^{1/2} \mbf M_i \mbf \Sigma_{H_0}^{1/2} $, and $\lambda_j$s are the eigenvalue of the constant matrix $\mbf \Tilde{M}$.

Based on equation \ref{eq:Quadraticforms}, we construct a statistic, $T$, to test the presence of overall effects apart from the noise. In this context, the null hypothesis is
\begin{align}
\sigma^2_1=\sigma^2_2=\dots=\sigma^2_J=0.
\end{align}
With the null hypothesis, we formulate a statistic $T$ as
\begin{align}
T=\frac{\sum_{i=1}^J w_i \hat{\sigma}_i^2 }{\hat{\sigma}_J^2 },
\end{align}
where $w_i$ is a predefined weight for the $i$th variance component, often set to 1 for all components.

We further explore the distribution of the statistic $T$, expressed as
\begin{align}
P\left(\frac{\sum_{i=1}^J w_i \hat{\sigma}_i^2}{ \hat{\sigma}_0^2} <t\right) = P\left( \frac{\sum_{i=1}^J w_i \hat{\sigma}_i^2 -t \hat{\sigma}_0^2}{\sigma_0^2}<0\right)
\end{align}
where $T$ is viewed as a random variable, and $t$ comes from the MINQUE estimation.

Subsequently, we can write
\begin{align}
   \frac{\sum_{i=1}^J   w_i \hat{\sigma}_i^2 -t \hat{\sigma}_0^2}{\sigma_0^2}&= \frac{1}{\sigma_0^2} \sum_{i=1}^J   w_i  \mbf{\epsilon}^T\mbf M_i \mbf{\epsilon} - t \mbf{\epsilon}^T\mbf M_0 \mbf{\epsilon} \\
   &= \frac{1}{\sigma_0^2} \mbf{\epsilon}^T \left( \sum_{i=1}^J   w_i  \mbf M_i  - t \mbf M_0\right) \mbf{\epsilon}\\
   &=\frac{1}{\sigma_0^2} \mbf{\epsilon}^T \mbf M_v \mbf{\epsilon}
\end{align}
Under the null hypothesis, $\mbf  \epsilon \sim N(\mbf 0, \sigma_0^2 \mbf I)$, \begin{align}
    \frac{\sum_{i=1}^J   w_i \hat{\sigma}_i^2 -t \hat{\sigma}_0^2}{\sigma_0^2} \sim  \sum_j \lambda_j \chi_1^2
\end{align}
which follows a  mixture of chi-square distributions, and $\lambda_j$s are the eigenvalue of the  matrix $\mbf M_v$. 

Therefore, the probability $P\left(\frac{\sum_{i=1}^J w_i \hat{\sigma}_i^2}{ \hat{\sigma}_0^2} <t\right)$ is equivalent to the value of the cumulative distribution function (CDF) of a mixture of chi-square distributions evaluated at point 0.

\subsection{An Association Test Based on Kernel Neural Networks}
In the prior section, we formulated statistical tests for both individual parameters and the collective effect. In the context of our kernel-based neural network model framework, we utilize MINQUE to estimate the parameters $\theta_0,\theta_1,\theta_2,\theta_3$. As a result, these statistical tests can be used to evaluate the linear effect ($H_0: \theta_2=0$), the non-linear effect ($H_0: \theta_3=0$), as well as the overall effect ($H_0: \theta_2=\theta_3=0$).

In the single parameter testing, the $\mbf \Sigma_{H_0}$ could be replaced by the $\hat{\mbf \Sigma}_{H_0}$, which is estimated through a reduced model.  For example, to test the linear effect, the statistics will be 
\begin{align}
     \hat{\theta}_2^2 &= \mbf \epsilon^*{}' \Tilde{\mbf M}_2 \mbf \epsilon^{*} 
\end{align}
where $\Tilde{\mbf M}_2 = \hat{\mbf \Sigma}_{H_0}^{1/2} \mbf M_2 \hat{\mbf \Sigma}_{H_0}^{1/2}$ and  $\hat{\mbf \Sigma}_{H_0}$ is estimated from a reduced model 
$$
 \mrm{Var}[\mbf{y}] =\theta_1 \mbf{J}+\theta_3 \frac{1}{p^2}(\mbf{XX}^T)^{\owedge2}+\theta_0 \mbf{I}_n.
$$ A similar statistic could be constructed for testing the non-linear effect $\theta_3$.

In the overall effect testing, the underlying hypothesis is $\theta_1=0$ if $\theta_2=\theta_3=0$. Therefore, the statistics $T$ for the overall effect could be presented as
\begin{align}
    T=\frac{\sum_{i=2}^3 w_i \hat{\theta}_i^2}{ \hat{\theta}_0^2},
\end{align}
and the p-value of the statistic is defined as
\begin{align}
    P\left(\frac{\sum_{i=2}^3 w_i \hat{\theta}_i^2}{ \hat{\theta}_0^2} >t\right)=  P\left(\frac{ w_2\hat{\theta}_2 +w_3 \hat{\theta}_3^2 -\hat{\theta}_0^2 t }{ \theta_0} >0\right)
\end{align}
which follows a mixture of chi-square distribution as outlined in the previous section. Several approximations and exact methods have been proposed for determining the distribution of a mixture chi-square distribution \cite{duchesne2010computing}. Among these methods, the Davies exact method \cite{davies1980algorithm} seems to work effectively in practice and is employed here.

\subsection{Simulation}
In this section,  we conducted some simulations to compare the performance of the proposed method based on the kernel-based neural network model to the performance of the sequence kernel association test (SKAT) in the linear mixed model. All the simulations are based on 1000 individuals with 1000 Monte Carlo iterations. Simulations were conducted utilizing genuine genotype data from the UKB. The methodologies for quality control (QC) and a comprehensive description of the data are elaborated in \cite{bycroft2018uk}. Post-QC, 320,021 samples had been genotyped via the UK Biobank Axiom array and the UK BiLEVE Axiom array, culminating in 516,429 retained autosomal SNPs. For simplification, a subset of 61,463 samples was randomly chosen, excluding chromosome 6 due to the intricate architecture of the major histocompatibility complex (MHC) region. Consequently, the simulations were performed using the resultant 61,463 samples and 341,545 SNPs.

\subsubsection{Simulation I}

In simulation study I, we examine the performances of both methods under the situation of nonlinear random effects. Specifically, we used the following model to simulate the response:
\begin{align}
    \mbf{y}=f(\mbf{a}) + \mbf{\epsilon}, ,\quad\mbf{a}\sim\mcal{N}_n\left(\mbf{0},\sigma_g^2 \frac{1}{p_{\lambda}}\Tilde{\mbf{G}}  \Tilde{\mbf{G}}^T\right),
\end{align}
where $\Tilde{\mbf{G}}$ is a $n\times p_\lambda$  the casual genotype matrix which is a small proportion $\lambda$ of $n\times p$ SNPs matrix $\mbf{G}$; and $\mbf{\epsilon}\sim\mrm{ i.i.d. }\mcal{N}_n(\mbf{0},\sigma^2_0\mbf{I}_n)$. The trait was simulated under the null model or one of four types of functions $f$: e linear ($f(x)=x$), quadratic ($f(x)=x^2$), hyperbolic cosine ($f(x)=cosh(x)$) and ricker curve ($f(a)=\beta*r(a^2)exp(-r(a^2))$, where $r$ is the soft rectifier, $r(a)=log(1+e^a)$), respectively, to assess the type I error or the statistical powers.  The phenotype of 1000 individuals was generated based on 1,50, 500, and 4,000 SNPs randomly selected from the UKB under the null model or one of the nonlinear models.  The proportion of causal SNPs was fixed at 20\%.  Table \ref{tab:simIset} presents the parameter settings for all scenarios in Simulation I, where $\sigma_g^2$ represents the genetic effect, $\sigma_0^2$ denotes the noise, and $\beta$ is a parameter specific to the Ricker curve model. Each simulation scenario was replicated 1,000 times.

\begin{table}[htb]
\centering
\caption{The scenario settings in simulation I}
\label{tab:simIset}
\resizebox{\columnwidth}{!}{%
\begin{tabular}{ccccccccccccc}
\hline
                              & \multicolumn{3}{c}{1}                                                                       & \multicolumn{3}{c}{50}                                                                       & \multicolumn{3}{c}{500}                                                                      & \multicolumn{3}{c}{4000}                                                                      \\ \cline{2-13} 
\multirow{-2}{*}{Sample Size}& $\sigma_g^2$ & $\sigma^2_0$ & $\beta$ & $\sigma_g^2$ & $\sigma^2_0$ & $\beta$ & $\sigma_g^2$ & $\sigma^2_0$ & $\beta$ &  $\sigma_g^2$ & $\sigma^2_0$ & $\beta$  \\ \hline
Linear                        &0.30 &2.00 &1.00  &0.30  &2.00 &1.00  &0.30  &2.00 &1.00  &1.00  &2.00 &1.00   \\
Quadratic                        &0.50 &2.00 &1.00  &0.50  &2.00 &1.00  &2.00  &2.00 &1.00  &2.00  &2.00 &1.00   \\
Cosh                          &0.80 &2.00 &1.00  &0.80  &2.00 &1.00  &1.20  &2.00 &1.00  &1.50  &2.00 &1.00   \\
Rickercurve                   &0.50 &2.00 &30.00 &0.50  &2.00 &30.00 &0.50  &2.00 &70.00 &0.50  &2.00 &150.00 \\
Multiplicative   interaction (2- way )                     &NA   &NA   &NA    &1.50  &2.00 &1.00  &10.00 &2.00 &1.00  &20.00 &2.00 &1.00   \\
Multiplicative   interaction (3- way )                   &NA   &NA   &NA    &10.00 &2.00 &1.00  &20.00 &2.00 &1.00  &20.00 &2.00 &1.00   \\
Threshold   interaction (2-way)  &NA   &NA   &NA    &0.80  &2.00 &1.00  &3.00  &2.00 &1.00  &20.00 &2.00 &1.00   \\
Threshold   interaction (3-way)               &NA   &NA   &NA    &1.00  &2.00 &1.00  &5.00  &2.00 &1.00  &20.00 &2.00 &1.00   \\ \hline
\end{tabular}
}
\end{table}

In Table \ref{tab:TypeIerror}, we present the  Type I error rates of our proposed method for continuous phenotypes across various null model scenarios, with a significance threshold set at $\alpha=0.05$. Our findings emphasize the robustness of our method in protecting against Type I errors. It's worth noting that in cases with smaller sample sizes, the SKAT method shows a conservative bias. Most observed Type I error rates align closely with their expected levels, even in some extreme scenarios, such as a single SNP. Specifically, for SNP counts of 1, 50, 500, and 4000, our method yielded overall Type I error rates of 0.044, 0.043, 0.059, and 0.043, respectively. When broken down by effect type, linear effects showed rates of 0.041, 0.044, 0.050, and 0.053, while the non-linear effects reported 0.050, 0.050, 0.052, and 0.057. On the other hand, the SKAT method exhibits strong conservativeness, especially in high-dimensional data analysis. With just 1 SNP, SKAT's Type I error rate was a notably conservative 0.024. While the error rate increased to 0.040 at an SNP count of 50, it sharply dropped as the SNP counts grew,  0.036 for 500 SNPs and 0.000 for 4000 SNPs. In conclusion, these empirical observations support the validity and efficiency of the MINQUE testing approach, especially when compared to current methods in high-dimensional data situations.

\begin{table}[htbp]
\centering
\caption{The Type I Error comparison between  kernel-based neural network model  Testing and SKAT under significant level 0.05}
\label{tab:TypeIerror}
\begin{tabularx}{\textwidth}{ X  X  X  X X }
\hline
\textbf{SNP} & \textbf{Total} & \textbf{Linear} & \textbf{Non-linear} & \textbf{SKAT} \\ \hline
        1 & 0.044 & 0.041 & 0.050 & 0.024 \\
        50 & 0.043 & 0.044 & 0.050 & 0.040\\
        500 & 0.059 & 0.050 & 0.052 & 0.036 \\ 
        4000 & 0.043 & 0.053 & 0.057 & 0.000\\ \hline
\end{tabularx}
\end{table}

\begin{figure}[htb]
\noindent\includegraphics[width=\linewidth]{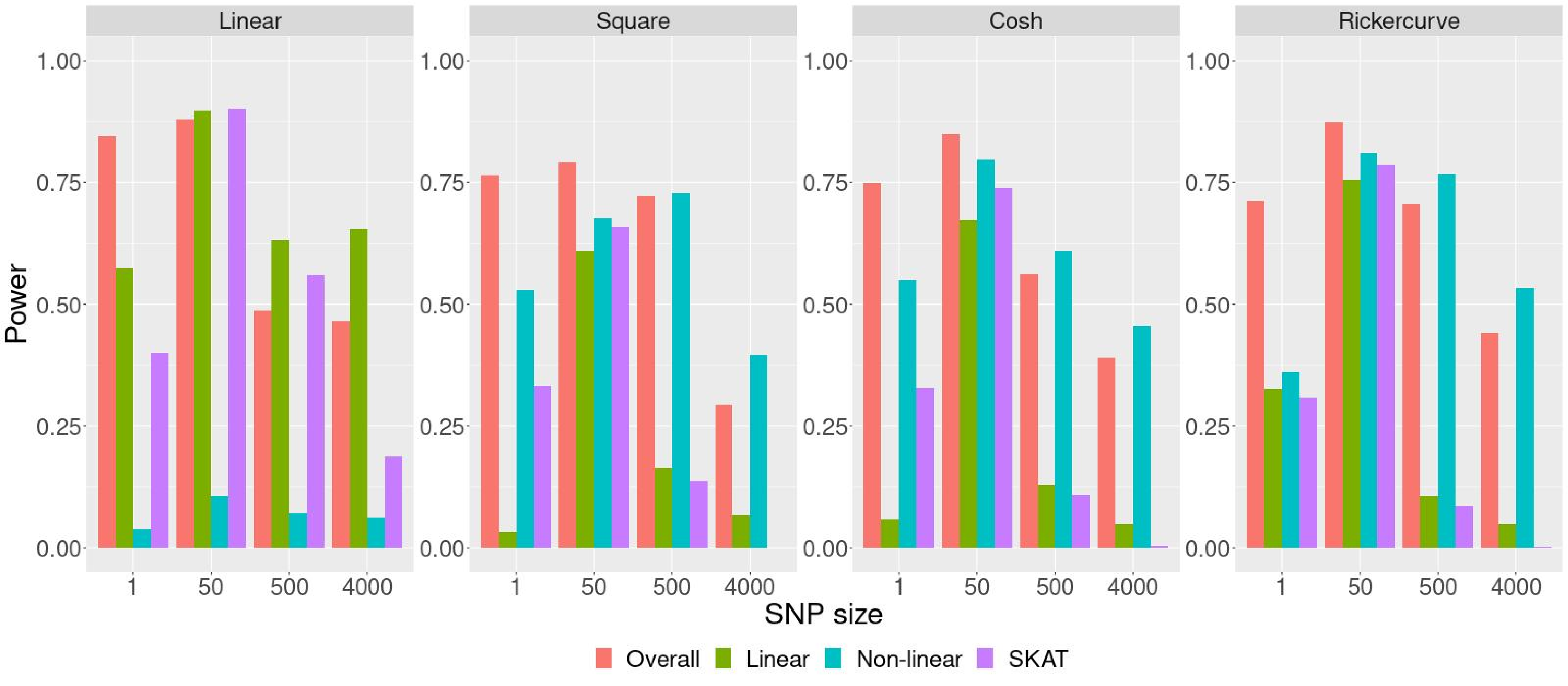}
  \caption{Comparative Power Analysis of Our Method and SKAT Across Diverse SNP Counts in Linear and Non-linear Simulations. Panel titles indicate specific simulation scenarios. The horizontal axis enumerates the SNPs involved in each simulation. Red Bar – Overall genetic effect of our method; Green Bar – Linear genetic effect of our method; Blue Bar – Nonlinear genetic effect of our method; Purple Bar – SKAT method's power for reference.}
  \label{fig:nonlinear}
\end{figure}

Figure \ref{fig:nonlinear} compares our method with the SKAT across different SNP sizes in both linear and non-linear settings. In the linear scenarios, the statistical power of our method is closely matched with the SKAT. Specifically, when considering extreme situations, our method shows a distinct advantage. At a single SNP (SNP=1), the power of our method stands at 0.846, whereas the SKAT registers a significantly lower 0.4. As we expand the analysis to larger SNP sizes, the difference in power between the two methods starts to narrow. For instance, with 50 SNPs, our method achieves a power of 0.88, while the SKAT is a slightly higher 0.901. At the 500 SNPs, our method's power is 0.487, with the SKAT measuring at 0.559. The advantage of our method becomes especially evident in high-dimensional situations. When analyzing a massive 4000 SNPs dataset, our method's power is 0.464, which is much higher than the SKAT's 0.062.

In non-linear scenarios, our method consistently demonstrates superior capability in distinguishing between linear and non-linear effects compared to the SKAT. Across the square, Cosh, and Rickercurve models with SNP counts of 1, 50, 500, and 4000, our method consistently outshines SKAT in overall effect power. For instance, in the square model at SNP=1, our method exhibits a power of 0.764 compared to SKAT's 0.332, and the difference becomes even more significant at SNP=4000, where our method achieved a power of 0.294 compared to SKAT's mere 0.001. The Cosh model shows a similar trend; our method achieves a power of 0.749 at SNP=1, overshadowing SKAT's 0.328, and maintaining the lead with 0.391 to SKAT's 0.004 at SNP=4000. In the Rickercurve model, our method's dominance persists, posting a 0.712 power at SNP=1 against SKAT's 0.309, and widening the gap with 0.44 against SKAT's 0.002 at SNP=4000. 
This consistent superiority underscores our method's robustness and precision in a diverse range of analytical settings.

Furthermore, the simulation study demonstrated the remarkable capability of our method to distinguish whether genetic variants had a linear or nonlinear influence on complex diseases. In the "square" model at SNP=500, our model's capabilities are starkly highlighted. The power of our method to detect the overall genetic effect stood strong, but it's crucial to dissect where this power primarily stems from. The linear genetic effect showcased a power of 0.163, whereas the nonlinear genetic effect was considerably more robust, with a power of 0.728. This significant disparity in power between the two effects points towards the nonlinear effect being the predominant contributor to the overall genetic effect detected by our model. It underscores the method's finesse in not only detecting genetic effects but also distinguishing between linear and nonlinear impacts.

 \clearpage
\subsubsection{Simulation II}
In this simulation study, we explore the performances of both methods under non-additive effects. In simulation studies, we mainly focus on the interaction effect and generate the response using the following model:
$$
\mbf{y}=f(\Tilde{\mbf{G}})+\mbf{\epsilon},
$$
where $\Tilde{\mbf{G}}=[\mbf{g}_1,\ldots,\mbf{g}_{ p_\lambda}]\in\mbb{R}^{n\times  p_\lambda}$ is the SNP data and $\mbf{\epsilon}\sim\mcal{N}_n(\mbf{0},\mbf{I}_n)$. When applying both methods, the mean is adjusted so that the response has a marginal mean of 0. In the simulation, we considered two types of interaction models, the multiplicative interaction and the threshold interaction,  with the following functions.

\begin{figure}[htb]
\noindent\includegraphics[width=\linewidth]{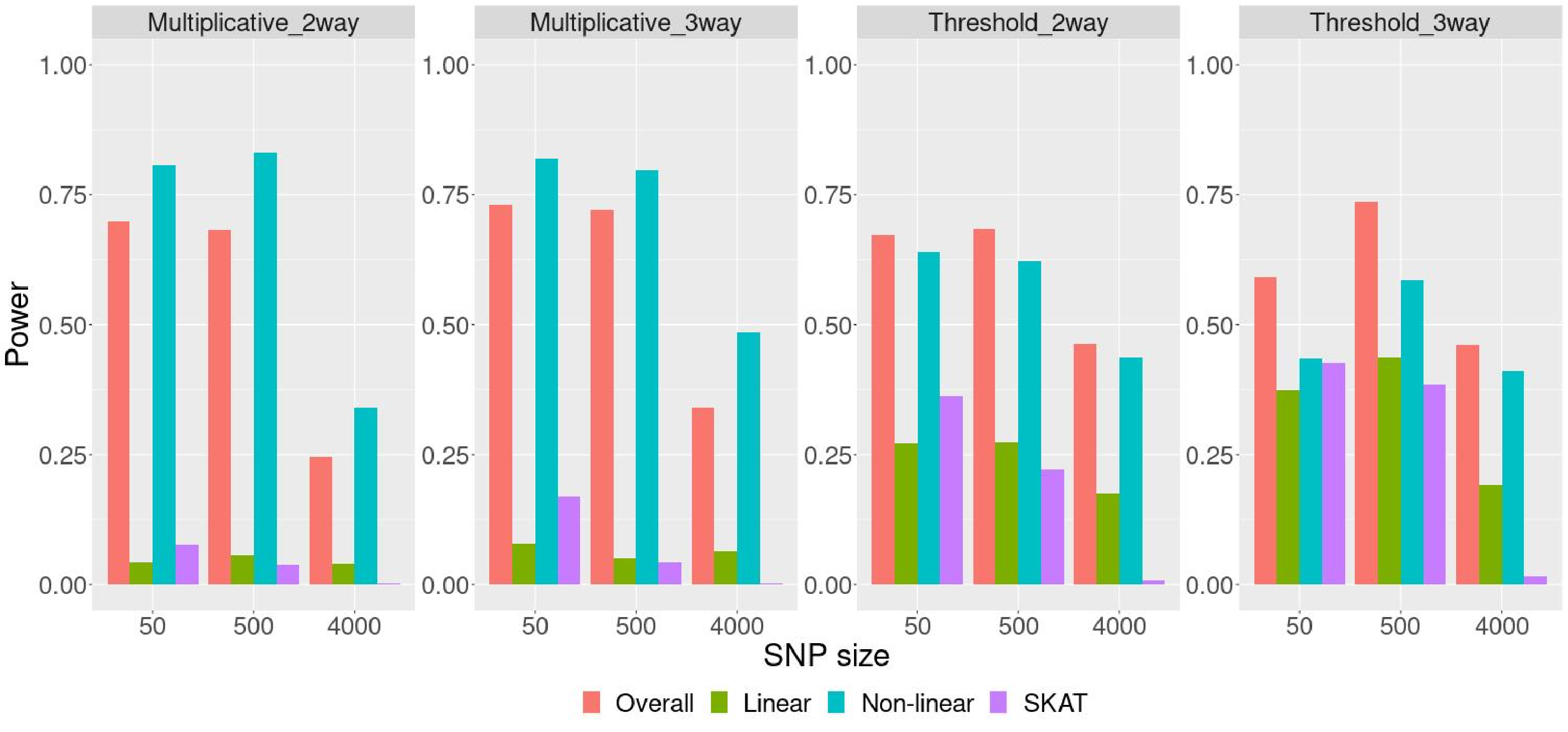}
  \caption{The Comparative Power Analysis of Our Method and SKAT Under Multiple Interaction Models. Panel titles indicate specific simulation scenarios. The horizontal axis enumerates the SNPs involved in each simulation. Red Bar – Overall genetic effect of our method; Green Bar – Linear genetic effect of our method; Blue Bar – Nonlinear genetic effect of our method; Purple Bar – SKAT method's power for reference.}
  \label{fig:NonAdditive}
\end{figure}

The multiplicative interaction
\begin{align}
    f(\Tilde{\mbf{G}})=\sum_{1\leq j_1<j_2\leq p_\lambda }\mbf{g}_{i_{j_1}}\odot\mbf{g}_{i_{j_2}}
\end{align}
The threshold interaction
\begin{align}
f(\Tilde{\mbf{G}})= \begin{cases}
    \sum_{1\leq j_1<j_2\leq p_\lambda }\mbf{g}_{i_{j_1}}\odot\mbf{g}_{i_{j_2}}, &\quad \text{if}\quad \sum_{1\leq j_1<j_2\leq p_\lambda }\mbf{g}_{i_{j_1}}\odot\mbf{g}_{i_{j_2}} >0 \\
    0,  &\quad \text{otherwise.}
\end{cases}
\end{align}

where $\odot$ stands for the Hadamard product. The phenotype of 1000 individuals was generated based on 1,50, 500, and 4,000 SNPs randomly selected from the UKB under the one of non-additive models.  The proportion of causal SNPs was fixed at 20\%.  Table \ref{tab:simIset} presents the parameter settings for all scenarios in Simulation II, 

Figure \ref{fig:NonAdditive} presents simulation outcomes across several interaction models. In every scenario, our method consistently outperformed SKAT. Within the 2-way multiplicative interaction framework, our method achieved powers of 0.699, 0.682, and 0.246 for 50, 500, and 4000 SNPs, respectively. In contrast, SKAT displayed powers of 0.077, 0.038, and 0.002 for the same SNP counts. Similarly, in the 2-way threshold interaction model, our method's powers for 50, 500, and 4000 SNPs were 0.673, 0.684, and 0.462, whereas SKAT's were 0.363, 0.221, and 0.009. This trend of our method's superior performance persisted in the 3-way interaction models, both multiplicative and threshold.  For the  3-way multiplicative model with SNP sizes of 50, 500, and 4000, our method achieves overall powers of 0.731, 0.721, and 0.34, respectively. In contrast, SKAT reports powers of 0.169, 0.043, and 0.002 for the same SNP sizes. Similarly, in the 3-way threshold model, our method exhibits overall powers of 0.592, 0.737, and 0.46 for SNP sizes of 50, 500, and 4000, respectively. Meanwhile, SKAT lags behind with powers of 0.427, 0.385, and 0.016 for these SNP sizes. These results consistently highlight the robustness of our model across diverse SNP sizes, effectively capturing complex interaction effects and demonstrating a distinct edge over existing techniques like SKAT.

Additionally, our model adeptly discerned if genetic variants had a linear or nonaddictive impact on complex diseases based on the kernel-based neural network model framework. For instance, with a set of 50 SNPs, our model exhibited an overall power of 0.699. But when we dissected the contributions of linear versus nonlinear effects, a significant disparity emerged: the linear effect, often considered a more straightforward genetic association, only achieved a power of 0.043. On the other hand, the nonlinear effect, which captures complex gene-gene interactions, showcased a remarkable power of 0.808. For comparison, using the SKAT method yielded a power of only 0.077 for the same set of 50 SNPs. Similar trends were found in other interaction models. Similar trends were observed in other interaction models. These results demonstrate that our model consistently excelled at identifying and emphasizing these nonlinear relationships, underscoring its proficiency in decoding complex genetic interactions that play a crucial role in complex diseases.

\subsubsection{Simulation III}
In this simulation study, we conducted an in-depth analysis to understand the influence of diverse weight functions and varying proportions of causal SNPs on performance outcomes. The primary goal was to compare the statistical power of our proposed method with the SKAT. By systematically examining these parameters, we aimed to shed light on how different combinations of weight functions and SNP causality proportions impacted the results. This comparative analysis offers vital insights into the robustness and potential constraints of our method in various scenarios. Specifically, our simulations involved 500 SNPs and generated data for 1,000 individuals with SNP causality proportions set at 20\% and 80\% across different models, including linear/nonlinear and non-additive models. Moreover, we incorporated two distinct weight functions in our study: unweighted (UW) and Beta function (BETA). The detailed settings for this simulation are presented in \ref{tab:simIset}.

\begin{figure}[htb]
\includegraphics[width=\linewidth]{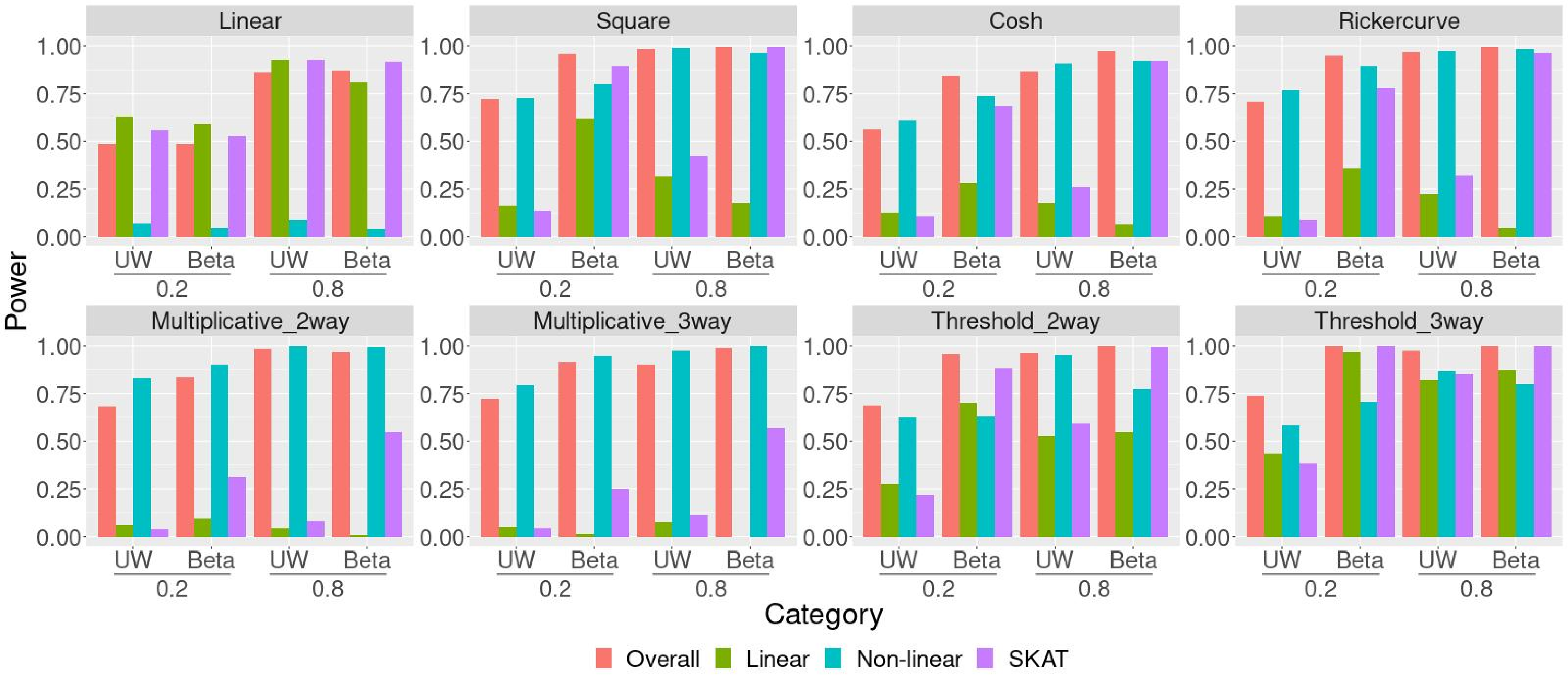}
  \caption{The Comparative Power Analysis of Our Method and SKAT Under diverse weight functions and varying proportions of causal SNPs. Each panel title signifies a distinct simulation scenario. The horizontal axis displays the weight functions used in the simulation, with "UW" representing the unweighted function and "Beta" denoting the beta function. Below these functions, numbers indicate the proportion of causal SNPs in each scenario. Red Bar – Overall genetic effect of our method; Green Bar – Linear genetic effect of our method; Blue Bar – Nonlinear genetic effect of our method; Purple Bar – SKAT method's power for reference.}
  \label{fig:WeigthRatio}
\end{figure}

Figure \ref{fig:WeigthRatio} shows the results of the simulation study, which consistently demonstrated the robust performance of our proposed method across different scenarios. Taking nonlinear models as examples, the advantages of our method become particularly evident. For instance, in the square model, with a proportion of causal SNPs set at 0.8, our method achieves a power of 0.987 in detecting overall genetic effects, significantly outpacing SKAT's 0.425. Even when the proportion of causal SNPs is reduced to 0.2, our method remains strong with a power of 0.722, far surpassing SKAT's 0.137. The story is consistent in the cosh model: with causal SNPs at 0.8, our method reaches a power of 0.866, overshadowing SKAT's 0.260. At the 0.2 mark, our method posts a power of 0.561, while SKAT trails at 0.108. Lastly, under the Rickercurve scenario with a 0.8 proportion of causal SNPs, our method impressively records a power of 0.97, easily besting SKAT's 0.322. With SNPs at 0.2, the figures stand at 0.706 for our method against SKAT's 0.086.  Similar trends appeared in the nonadditive model. 

Furthermore, the simulation results also demonstrated the robustness of our proposed method regardless of the choice of the weight functions. Take the square scenario as an example and focus on a 0.2 proportion of causal SNPs: when using the UW weight function, our method achieves a power of 0.722, substantially outperforming SKAT's 0.137. For the same scenario, but using the Beta weight function, our method maintains a robust power of 0.959, while SKAT lags behind at  0.890. At the 0.8 proportion level, with the UW function, our method records a power of 0.987 against SKAT's  0.425, and with Beta, our method stands at 0.996 compared to SKAT's 0.993.  In the given scenario, our method demonstrated a modest power disparity of only 0.237 between "UW" and "Beta" when the proportion of causal SNPs was at 0.2. This difference further diminished to a mere 0.009 when the proportion of causal SNPs increased to 0.8. Conversely, SKAT displayed a more significant fluctuation, specifically 0.753 and 0.568 for proportions of causal SNPs at 0.2 and 0.8, respectively.

These findings emphasize the effectiveness of our method in identifying overall genetic influences. Regardless of changes in weight functions or fluctuations in the proportion of causal SNPs, our method remains steadfastly reliable. 

\clearpage
\subsection{Real Data Application}
To showcase the capabilities of our proposed method, we applied it to the hippocampal volume data from the UK Biobank. This volume is derived from the average of the right and left hippocampal volumes. The UK Biobank is a comprehensive, detailed prospective study that includes around 500,000 participants aged 40-69 years, collected from 2006-2010 across the UK \cite{sudlow2015uk}. This study amassed a wide array of phenotypic and health data for each participant, ranging from biological measurements, lifestyle markers, and biomarkers in blood and urine, to body and brain imaging \cite{bycroft2018uk}. More information about the UK Biobank project can be accessed at \url{https://www.ukbiobank.ac.uk/}.

Within the UK Biobank, a total of 502,591 samples were collected, and genotyping of over 800,000 markers was carried out using two arrays, both with over 95\% shared marker content - namely the UK Biobank Axiom array and the UK BiLEVE Axiom array. These markers include both SNPs and small insertions and deletions (Indels) \cite{bycroft2017genome}. The quality assurance measures followed were in line with prior GWAS studies. Specific criteria were set for the quality control procedure, including (a) participant call rate above 95\%; (b) SNP call rate above 99\%, only including SNPs from autosomes; (c) heterozygosity below 19.03\%; (d) relative kinship coefficient under 0.0442, excluding any individual from sample pairs with 3rd-degree or closer family ties; (e) SNPs with a minor allele frequency greater than 0.01 and HWE p-value above 1e-7; and (f) a participant's self-reported race and ethnicity matching the findings from PCA, narrowing our analysis to those identifying as white British and verified by principal component analysis (PCA) as Caucasians. After these stringent quality checks, data from 320,021 participants and 516,429 potential SNPs were deemed suitable for kernel-based neural network model analysis. All these quality assurance procedures were conducted using PLINK 1.9 (http://www.cog-genomics.org/plink2) \cite{chang2015second}. We tested 516,429 SNPs for their association with the hippocampus volume and assigned them to each of the 24,219 autosomal genes based on their positions on the human assembly GRCh37 (also known as hg19) in Ensembl (\url{https://useast.ensembl.org/index.html}). To capture regulatory regions and SNPs in linkage disequilibrium, we defined gene boundaries as $\pm$ 5 kb of 5' and 3' UTRs. Normalized product kernel matrices were constructed for each gene for analysis. We also considered age, gender, race, education, APOE4, and the top 10 PCs as covariates.

The table \ref{tab:MixtureChiRealData} shows the top 10 significant genes associated with the hippocampal volume as identified by the mixture chi-square test in UK Biobank data.  With the results, we observed substantial similarities and differences between our proposed method and SKAT in identifying genes associated with hippocampal volume. The most significant genes of our proposed method are overlapped with the results from SKAT. Taking the gene TFDP2, the most significant gene in our results, as an example, our method identified a significant overall effect with a p-value of 0.00002, and within this, a discernible linear effect was observed at a p-value of 0.00116 while the nonlinear effect wasn't significant. In comparison, SKAT's analysis highlighted a strong linear effect with a p-value of 0.00001. Other genes, such as SHMT2, NXPH4, and PTCSC2, were significant in our study, presenting overall genetic effect p-values of 0.00024, 0.00024, and 0.00042, respectively. Within these, the linear effect p-values stood at 0.00052, 0.00052, and 0.00213, with no notable nonlinear effects. Comparatively, SKAT's results for these genes were 0.00009, 0.00009, and 0.00034, respectively.

However, when evaluating nonlinear effects, the distinctive strength of our method becomes evident. Consider the gene LOC105372440. While SKAT presented a p-value of 0.04047, suggesting non-significance, our analysis emphasized its significant nonlinear contribution. Specifically, the gene had an overall genetic effect p-value of 0.00016. Though the linear effect appeared non-significant at a p-value of 0.18467, the nonlinear effect stood out with a significant p-value of 0.00003.  This suggests that LOC105372440 might have a nonlinear effect on the hippocampus volume rather than a linear effect. In contrast, SKAT indicated only a marginal significance with its p-value of 0.04047. Similarly, the gene AMMECR1L revealed a significant nonlinear genetic effect in our analysis, showing an overall effect p-value of 0.00024 and a nonlinear effect p-value of 0.00027, while SKAT reported a p-value of 0.02521.

\begin{table}[htb]
\centering
\caption{Top 10 significant genes associated with hippocampus volume using mixture chi-square test}
\label{tab:MixtureChiRealData}
\begin{tabularx}{\textwidth}{X  l   X   X   X   X}
\hline
CHR & GEN & Total & Theta1 & Theta2 & SKAT \\ \hline
3 & TFDP2 & 0.00002 & 0.00116 & 0.52922 & 0.00001 \\
19 & SMIM47 (LOC105372440) & 0.00016 & 0.18467 & 0.00003 & 0.04047 \\
12 & SHMT2 & 0.00024 & 0.00052 & 0.41070 & 0.00009 \\
2 & AMMECR1L & 0.00024 & 0.32782 & 0.00027 & 0.02521 \\
12 & NXPH4 & 0.00024 & 0.00052 & 0.51888 & 0.00009 \\
2 & POLR2D & 0.00033 & 0.06058 & 0.05888 & 0.00184 \\
9 & PTCSC2 & 0.00042 & 0.00213 & 0.19205 & 0.00034 \\
4 & TMEM150C & 0.00045 & 0.02394 & 0.12113 & 0.00096 \\
5 & TMEM174 & 0.00047 & 0.06739 & 0.30466 & 0.00040 \\
8 & MAL2-AS1 & 0.00068 & 0.00497 & 0.69399 & 0.00031 \\ \hline
\end{tabularx}%
\end{table}

\clearpage
\section{Discussion}
\label{sec:discussion}

In this study, we proposed an association testing approach based on the kernel-based neural network framework, designed to evaluate the joint association between multiple genetic variants and complex diseases. A standout feature of our methodology, as contrasted with our prior work and other prevalent methods like the SKAT, is its reliance on a mixture of chi-square distribution. This approach refrains from substituting estimated values for true ones, a practice that can lead to inflated or conservative type I errors in other techniques. Such substitutions have been observed in other methodologies and can inadvertently introduce inflated or conservative type I errors, muddying the clarity and reliability of results. However, in our proposed method, the meticulous design ensures a more accurate and trustworthy assessment, especially for overall genetic effect testing. Furthermore, our proposed Wald-type test provides a comprehensive evaluation of the links between genes and intricate diseases by accounting for non-linear and non-additive effects, which are often missed in conventional genetic association studies.

Moreover, our proposed method simplifies the computational process, making it both straightforward and efficient. It primarily hinges on the initial value of the estimation and the kernel matrices. This is in sharp contrast to more cumbersome methods like the normal kernel-based neural network testing, where the iterative MINQUE procedure is employed to calculate the estimated variance matrix of the estimators—a process that is both time-consuming and computationally taxing. Our approach not only streamlines this, but also significantly reduces the computational load. Additionally, the traditional MINUQE method, recognized for its complexity, can be seamlessly replaced by the more streamlined MINQUE(0). This alternative uses an initial variance component estimation in which all initial guesses, excluding the error component, are set to zero. This further reduces computational demands, allowing for a more efficient use of resources. The inherent efficiencies and accuracies of our method position it as a potent tool for genetic researchers, enabling in-depth exploration without the encumbrances of traditional computational constraints.

The \textit{TFDP2} gene, also known as Transcription Factor Dp-2, is a member of the transcription factor DP family. This gene encodes a protein that forms heterodimers with the E2F transcription factors, which are involved in the control of cell-cycle progression from G1 to S phase. Although the specific relationship between the  \textit{TFDP2} gene and the hippocampus volume is not well established, some Genome-Wide Association Studies (GWAS) identified some loci within the  \textit{TFDP2}  gene significantly associated with the volume of the hippocampus. A study involving brain scans from 21,297 individuals identified a significant locus with lead SNP rs7630893 at chromosome 3 within the TFDP2 gene\cite{van2020brain}.   Another study identified TFDP2 among several genes in the hippocampus\cite{ayhan2021resolving}. These studies suggest that variations in the \textit{TFDP2}  gene could potentially influence the structure of the hippocampus, although the exact mechanisms and implications of this association require further investigation. 
Another gene \textit{SMIM47},  is a protein-coding gene that is located on chromosome 19q13.33. Small Integral Membrane Protein 47   (SMIM47)  is an integral membrane protein, which may play a role in the movement of molecules across cell membranes \cite{whitelegge2013integral}. These processes are critical for the proper function of neurons in the hippocampus and other brain regions. The local translation of integral membrane proteins in the dendrites of hippocampal neurons has been shown to occur in response to plasticity-inducing patterns of activity \cite{grigston2005translation}. This suggests that integral membrane proteins may be involved in synaptic plasticity, which is the process by which synapses are strengthened or weakened in response to experience. All these studies show that the gene  \textit{SMIM47} may play an important role in hippocampus volume via some indirect pathway. 

The association between other genes and hippocampal volume remains uncertain, although there is evidence suggesting a possible connection with Alzheimer's disease. For example, the gene \textit{SHMT2} encodes for serine hydroxymethyltransferase,  a key enzyme in one-carbon metabolism, which can convert serine into a one-carbon unit \cite{ma2023vital}. One-carbon metabolism, also known as folate metabolism, plays a significant role in various cellular processes, including DNA methylation, synthesis, and repair\cite{garcia2020impairment}. It is required for the production of S-adenosylmethionine (SAM), which is the major DNA methylating agent\cite{coppede2010one}. Disruptions in one-carbon metabolism and elevated homocysteine have been previously implicated in the development of dementia associated with Alzheimer’s disease \cite{kalecky2022one}.In individuals with AD, decreased plasma folate values and increased plasma homocysteine (Hcy) levels have been observed, indicating impaired SAM levels in AD brains\cite{coppede2010one}.   DNA methylation alterations have been observed in various brain regions of individuals with AD \cite{yokoyama2017dna}. These alterations could be a result of disruptions in one-carbon metabolism, leading to changes in the methylation status of DNA, which can affect gene expression and contribute to the pathogenesis of AD \cite{yokoyama2017dna}. Therefore, the gene  \textit{SHMT2} may play an important role in the pathogenesis of Alzheimer's disease. In addition to SHMT2, the gene \textit{AMMECR1L}  was also reported to be associated with Alzheimer's disease \cite{hebert2023genetic}. Another gene \textit{NXPH4} is associated with cognitive impairment in Alzheimer's disease\cite{clark2021integrative}. Furthermore, \textit{POLR2D} is involved in the process of RNA splicing. Altered expression of RNA splicing proteins, including POLR2D, has been observed in Alzheimer's disease patients, suggesting that dysregulation of pre-mRNA splicing may underlie aberrant gene expression changes in Alzheimer's disease\cite{wong2013altered}. These studies show that genes are involved in the pathogenesis of Alzheimer's disease. However,  further research is needed to understand the potential roles of these genes.

In our research, we have presented a novel approach to association testing through the implementation of a kernel-based neural network. This method has demonstrated a remarkable increase in detection power over the traditional sequence kernel association test (SKAT), especially when addressing complex scenarios characterized by non-linear patterns and interactive effects between genetic variables. The enhanced capability of our method to capture intricate relationships in genetic data represents a significant advancement in the field of genetic epidemiology. It allows for a more nuanced understanding of how genes may interact with each other and with environmental factors to influence disease phenotypes. Our findings suggest that integrating neural network architectures with kernel methods can uncover subtle genetic associations that might otherwise be missed by more conventional approaches. This could be particularly beneficial in studying complex diseases like Alzheimer's, where the interplay of multiple genetic factors is not well explained by linear models. The potential of our approach to improve the accuracy of genetic association studies holds promise for the identification of novel risk factors and the development of personalized therapeutic strategies.

\bibliography{reference}
\end{document}